\documentclass[
preprintnumbers,tightlines,
amsmath,amssymb,aps,nofootinbib,nobibnotes]{revtex4}
\usepackage[colorlinks=true, pdfstartview=FitV, linkcolor=green, citecolor=blue, urlcolor=red]{hyperref}
\usepackage{amssymb}
\usepackage{latexsym}
\usepackage{epsfig}  
\usepackage{caption}
\usepackage{subcaption}
\usepackage[normalem]{ulem} 
\usepackage{color}
\usepackage{comment}
\usepackage{orcidlink}
\allowdisplaybreaks
\begin{document}
\title{Generalized First Law and Smarr-type Relation: Beyond Additivity and Extensivity}
\author{Usman Zafar $^{1}$ \orcidlink{0000-0001-9610-1081}\footnote{\color{red}s2471001@ipc.fukushima-u.ac.jp,~zafarusman494@gmail.com}, Krishnakanta Bhattacharya $^{2}$\orcidlink{0000-0003-3309-2610}
\footnote{\color{red}krishnakanta@dubai.bits-pilani.ac.in}, and Kazuharu Bamba$^{1}$\orcidlink{0000-0001-9720-8817}
\footnote{\color{red}bamba@sss.fukushima-u.ac.jp}}
\address{$^1$ Faculty of Symbiotic Systems Science, Fukushima University,
Fukushima 960-1296, Japan\\
$^2$ Department of General Science, Birla Institute of Technology and Science, Pilani, Dubai Campus, Dubai International Academic City, Dubai, United Arab Emirates}

\begin{abstract}
In this work, we develop a general framework for deriving the first law of black hole thermodynamics and the associated Smarr-type relation for 4-dimensional static spherically symmetric spacetimes within a broad class of generalized entropy models. In addition, by employing the entropy representation for the Reissner--Nordstr\"{o}m (RN) black hole, we determine a concise curvature functional that depends only on this representation and its derivatives. A general flatness condition is derived, demonstrating that when the entropy Hessian metric is regular and nondegenerate, both homogeneous power-law transformations and a separate logarithmic branch leave the RN black hole's Ruppeiner geometry flat. On the other hand, R\'{e}nyi, Sharma-Mittal, Kaniadakis, and LQG-inspired entropy deformations typically give rise to nonvanishing curvature in the regular thermodynamic regime, demonstrating that the extremality of the RN black hole does not by itself require a curvature divergence.
\end{abstract}

\maketitle

\section{Introduction}
Black hole (BH) thermodynamics has become a significant framework for exploring the intriguing interplay among gravity, statistical physics, and quantum mechanics. The profound connection between general relativity (GR) and thermodynamics, originally demonstrated by the groundbreaking work of Hawking and Bekenstein \cite{Hawking:1974rv,Bekenstein:1973ur,Bekenstein:1974ax,Hawking:1975vcx,Bekenstein:1972tm,Bardeen:1973gs}, continues to provide crucial insights that reinforce this foundational link. They suggested that BHs have entropy proportional to their event horizon area and emit thermal radiation, thereby making the thermodynamic interpretation of spacetime fundamental to quantum gravity research. The standard Boltzmann-Gibbs (BG) formalism for statistically independent subsystems becomes extensive under the traditional assumption of long- or short-range interactions \cite{Swendsen,Dabrowski:2024qkp}. The presence of long-range interactions and strong correlations in gravitational systems motivated the development of a generalized entropy framework that accounts for deviations from standard additivity and extensivity \cite{Nojiri:2022aof,Nojiri:2022dkr,Nojiri:2022sfd,Elizalde:2025iku,Odintsov:2023vpj,Nojiri:2024zdu,Odintsov:2022qnn,Tsallis:1987eu,Barrow:2020tzx,Rényi,SayahianJahromi:2018irq,Majhi:2017zao,Chatterjee:2020iuf,Calcagni:2026xld}. Since entropy reflects the microscopic degrees of freedom of BHs, any modification to the entropy formulation, whether arising from quantum gravity effects or generalized statistical approaches, directly affects the system's thermodynamic structure. Therefore, it is crucial to derive the first law (FL) and the Smarr relation consistently within each entropy model to maintain the internal consistency and physical validity of BH thermodynamics. Additionally, an intriguing question arises: is it possible to define the Smarr-type relation and the FL independently for the generalized entropy model within the class of four-dimensional spherically symmetric spacetimes (SSS) considered here?    

Another important feature of BH thermodynamics is the analysis of phase transitions, which can be classified into types such as Davies-type \cite{Davies:1989ey,Lousto:1994jd,Muniain:1995ih}, Hawking-Page \cite{Hawking:1982dh}, van der Waals-like \cite{Kastor:2009wy,Dolan:2010ha,Dolan:2011xt,Dolan:2011jm,Dolan:2012jh,Kubiznak:2016qmn,Bhattacharya:2017nru}, and extremal \cite{Pavon:1988in,Pavon:1991kh,Cai:1996df,Cai:1998ep,Wei:2009zzf,Bhattacharya:2019awq}. Thermodynamic geometry, first formulated by Weinhold \cite{weinhold1975metric} and subsequently developed by Ruppeiner \cite{Ruppeiner:1979bcp,ruppeiner1995riemannian,Ruppeiner:2008kd,Ruppeiner:2013yca}, provides a geometric framework for examining thermodynamic stability, critical behavior, and phase transitions. In BH thermodynamics, singularities of the Ricci scalar curvature are often associated with critical behavior or the divergence of thermodynamic response functions. However, the precise correspondence between thermodynamic curvature singularities and critical behavior depends on the choice of thermodynamic metric, potential and statistical ensemble \cite {Kubiznak:2016qmn,Bhattacharya:2017nru,Bhattacharya:2024bjp,Bhattacharya:2021lgk,Gunasekaran:2012dq,Hendi:2012um,Dolan:2013ft,Altamirano:2013ane,Chamblin:1999tk,Cvetic:1999rb,Wei:2017icx,Banerjee:2011au,Jawad:2020ihz,Myung:2006tg,Surya:2001vj,Guo:2019oad}. Various thermodynamic-geometric formalisms have therefore been employed to characterize thermodynamic behavior of the BHs under different entropy models. 

In this work, we address these questions by constructing a generalized framework that is independent of any particular BH solution or specific entropy model within the class considered here. In this regard, we first focus on Einstein's GR, where the thermodynamic FL and the Smarr-like relation are derived directly from Einstein's field equations for SSS \cite{Smarr:1972kt}. We construct the framework for the generalized first law and the Smarr relation for the class of four-dimensional static, spherically symmetric BH solutions. Furthermore, to make this generalized framework more rigorous, we formulate it to be independent of any particular entropy model by employing the generalized entropy. This generic formalism is valid for any static and spherically symmetric metric satisfying $g_{00}=-(g_{rr})^{-1}$, indicating that Einstein's field equation inherently comprises both the FL and the Smarr relation, and thus points towards a deeper thermodynamic interpretation of gravity consistent with its emergent nature. Motivated by the horizon-based construction developed in Ref.~\cite{Bhattacharya:2021lgk}, we adopt the corresponding geometric relation as the foundation for extending the FL and Smarr-type relation to generalized entropy frameworks. The present analysis is therefore distinguished not by a new derivation of the traditional horizon identity, but by its extension to a general entropy-functional framework and by an investigation of how homogeneous and non-homogeneous entropy models affect the corresponding thermodynamic relations. In the second part of our analysis, we extend the analysis to examine the extremal limit by developing a generalized Ruppeiner thermodynamic geometry formulated in terms of the generalized entropy. In this regard, we defined the Ruppeiner metric in the form of the Hessian of the generalized entropy (which incorporates the parameter characterizing deviations from the standard entropy) with respect to the relevant thermodynamic variables, thereby enabling one to write the Reissner-Nordstr\"{o}m (RN) thermodynamic metric and curvature to be expressed functionally in terms of the generalized entropy map. Then, we apply this framework to the outer horizon of the RN BH by employing intrinsic two-dimensional phase space. Utilizing this framework, one can obtain a generic relation for the thermodynamic curvature scalar, which enables the geometric structure of the thermodynamic phase space to be investigated without the need for any particular entropy model (numerous works have been done to explore different BH models and different entropy models by using thermodynamic geometry; for further details, see Refs.~\cite{Rani:2022xza,Wei:2012ui,Aman:2005xk,Soroushfar:2020wch,Xu:2020gud,Chaudhary:2022sfg,Zhang:2014uoa,Bravetti:2012dn,Mirza:2007ev,Jawad:2022lww,Zafar:2025nho}). Furthermore, the curvature functional indicates that RN extremality  does not inherently lead to a curvature divergence; rather, such behavior is governed by the entropy map and regularity of the Hessian metric.

We derive the general criterion governing thermodynamic flatness and explicitly demonstrate that, for a regular, nondegenerate metric, the flat Ruppeiner geometry of the RN BH is preserved under a homogeneous power-law entropy transformation, as well as a separate logarithmic branch. This paper is arranged as follows. We explicitly derive the generalized BH thermodynamics' FL and the Smarr relation from Einstein's field equations in
Sec.~\ref{sec-2}. Moreover, we derive the entropy functional Ruppeiner geometry of the RN BH and investigate its extremal limit in Sec.~\ref{sec-4}.  Lastly, Sec.~\ref{sec-6} presents a detailed analysis of our results.  In Appendices~\ref{app:A} and \ref{app:B}, we discuss the corresponding entropy limits and the basic construction of our Ruppeiner interpretation, respectively. 

\section{Generalized First Law of BH thermodynamics and the Smarr Formula}\label{sec-2} 
This section derives and discusses the thermodynamic FL and the generalized Smarr relation from the Einstein field equations. Evaluating the Einstein field equation on the horizon of the BH, one can obtain the thermodynamic FL of the BH as given in Refs.~\cite{Padmanabhan:2002sha,Padmanabhan:2003gd,Paranjape:2006ca,Kothawala:2007em,Hayward:1997jp,Bhattacharya:2016kbm,Hansen:2016gud,Bhattacharya:2021lgk,Battista:2020zxo,Battista:2022imy,Borsboom:2026ash,Mishra:2017sqs,Aneesh:2020fcr,Chakraborty:2015aja}. Throughout the manuscript, we adopt natural units such that $G_{\rm N}=\hbar=c=k_{\rm B}=1$. To begin with, we discuss the basic formalism by introducing a general SSS, which is given as 
\begin{eqnarray}\label{GST}
    ds^{2}=-f(r)dt^{2}+(f(r))^{-1}dr^{2}+r^{2}d{\tilde{\Omega}}^{2}\,,
\end{eqnarray}
where $d{\tilde{\Omega}}^{2}=d\theta^{2}+\sin^{2}\!\theta \  d\phi^{2}$ and $f(r)$ denotes the metric function of $r$. This metric function of $r$ fulfills the radial component of the Einstein field equation $G^{r}_{r}=8\pi T^{r}_{r}$, which yields
\begin{eqnarray}\label{EFR}
   \frac{f'(r)}{r}+ \frac{-1+f(r)}{r^{2}}=8\pi T^{r}_{r}\,.
\end{eqnarray}

In the literature, two approaches have been used to compute the thermodynamic FL described in Refs.~\cite{Padmanabhan:2002sha,Padmanabhan:2003gd}. In the first method, we utilize Eq.~\eqref{EFR} on the horizon radius $r=r_{\rm e}$, where $r_{\rm e}$ can be determined by setting the metric function of $r$ equal to zero. After multiplying both sides of Eq.~\eqref{EFR} by the term $r^{2}_{e}dr_{\rm e}/2$, we get 
\begin{eqnarray}\label{1WTL}
   TdS-dE=P(r)dV\,.
\end{eqnarray}
where $T$ is the conjugate temperature related to the entropy framework, $E$ represents the internal energy, which is obtained by interpreting the mass of the BH. The FL of BH thermodynamics is obtained from  Eq.~\eqref{1WTL} by taking infinitesimal displacement of the horizon from $r_{\rm e}$ to $r_{\rm e}+dr_{\rm e}$, which is given as 
\begin{eqnarray}\label{1WTL1}
   P_{(r)}=T^{r}_{r}|_{r_{\rm e}}\,,\quad V=\frac{4\pi r^{3}_{e}}{3}\,,\quad T=\frac{f'}{4\pi}\,,\quad S=\pi r^{2}\,, \quad E=\frac{r_{\rm e}}{2}\,.
\end{eqnarray}
Another approach to obtaining the thermodynamic law from Eq.~\eqref{EFR} is presented in the literature \cite{Hansen:2016gud}, as described below. In order to obtain the thermodynamic FL, we treat Eq.~\eqref{EFR} as the generalized equation of state as described in Refs.~\cite{Hansen:2016gud,Bhattacharya:2021lgk} and its explicit expression can be written as
\begin{eqnarray}\label{2WTL}
   P_{(r)}=B(r_{\rm e})+T C(r_{\rm e})\,,
\end{eqnarray}
where the expressions for $P(r)$ and $T$ are given in Eq.~\eqref{1WTL1}. Thereby, it is straightforward to obtain the thermodynamic relation by multiplying Eq.~\eqref{2WTL} with $V$ and applying the variation on each side, given as
\begin{eqnarray}\label{2WTL1}
   \delta P_{(r)}\ V=\delta G+S\delta T\,,
\end{eqnarray}
where 
\begin{eqnarray}\nonumber
  G&=&\int\limits^{r_{\rm e}}d\mathfrak{x}B'(\mathfrak{x}) V(\mathfrak{x})+T\int\limits^{r_{\rm e}}C'(\mathfrak{x})V(\mathfrak{x})d\mathfrak{x}\\\nonumber
  &=& V\ P_{(r)}-T S-\int\limits^{r_{\rm e}}B(\mathfrak{x})V'(\mathfrak{x})d\mathfrak{x}\,,\\\label{2WTL2}
  S&=&\int\limits^{r_{\rm e}}d\mathfrak{x}C(\mathfrak{x}) V'(\mathfrak{x})\,.
\end{eqnarray}
It is quite easy to find the values of $B(r_{\rm e})$ and $C(r_{\rm e})$ by comparing generalized equation of state \eqref{2WTL} with the Eq.~\eqref{EFR} which gives us $-\frac{1}{8\pi r^{2}_{e}}$ and $\frac{1}{2r_{\rm e}}$, respectively (for more details check Ref.~\cite{Bhattacharya:2021lgk}). So, by using the values of $B(r_{\rm e})$ and $C(r_{\rm e})$ in Eq.~\eqref{2WTL2}, we obtain the Gibbs free energy whose explicit form takes the following shape 
\begin{eqnarray}\label{2WTL3}
  G&=&\frac{r}{3}(1-T \pi  r_{\rm e})\,,
\end{eqnarray}
and the expression for $S$ is described in Eq.~\eqref{1WTL1}. Therefore, adopting the two distinct methods mentioned above yields the same thermodynamic quantities. By employing the Legendre transformation relation $E=TS+G-V\ P_{(r)}$, one can associate the internal energy provided in Eq.~\eqref{1WTL} and the Gibbs free energy given in the second method. 

To resolve the uncertainty in defining heat and work in the first method \cite{Padmanabhan:2002sha,Padmanabhan:2003gd}, the second method provides the entropy and volume, which depend solely on the horizon radius, as discussed in Refs.~\cite{Hansen:2016gud,Bhattacharya:2021lgk}. However, in both of these methods, additional care is required in identifying the corresponding thermodynamic potentials (for example, $E$ defined in the first method and $G$ in the second method) to match them with the relation that we computed from the FL of classical thermodynamics, and these potentials, which we obtained in both methods, are also not unique. For example, internal energy $E$ can be interpreted as the mass of the BH in the context of the Schwarzschild BH, but it cannot be the same in the case of the RN BH. Furthermore, if we incorporate the cosmological constant in the metric function, then it also becomes complicated in the extended phase space (for more details on the presence of cosmological constant, check Refs.~\cite{Kastor:2009wy,Dolan:2010ha,Kubiznak:2016qmn,Bhattacharya:2021lgk,Chamblin:1999tk,Gunasekaran:2012dq,Hendi:2012um}). The no-hair theorem further suggests that the parameters present in the BH metric are inherently more fundamental than the thermodynamic potentials introduced when applying the FL via these two methods. It can be noticed from Eq.~\eqref{1WTL} that the radial pressure is defined by $T_{r}^{r}(r_{\rm e})$. For the RN BH the radial pressure takes the following shape $p(r)=T_{r}^{r}=-\frac{Q^{2}}{8\pi r^{4}}-\frac{\Lambda}{8\pi}$ but for RN-AdS BH in AdS spacetime, we interpret thermodynamic pressure $P=-\frac{\Lambda}{8\pi}$ that plays a crucial role in the $P-V$ criticality, which also satisfies the extended phase space thermodynamics as discussed in Refs.~\cite{Kastor:2009wy,Dolan:2011xt,Bhattacharya:2017nru,Kubiznak:2012wp,Majhi:2016txt,Dolan:2013ft,Zafar:2025nho,Borissova:2026rbi}. This means the radial pressure should not be interpreted as the thermodynamic pressure when studying $P-V$ phase transitions in the extended phase space.

In addition, the laws of thermodynamics and the Smarr relation for BHs can be derived utilizing variables such as $M$ and $Q$, etc, without the need to define a thermodynamic potential, particularly focusing on whether variations in horizon area can be explained solely through changes in these parameters.  This formulation is obtained by solving the Einstein field equation. Therefore, one can obtain the thermodynamic FL and the Smarr relation by only utilizing the metric spacetime as discussed in Ref.~\cite{Bhattacharya:2021lgk}. An interesting question is raised regarding the role of entropy in these thermodynamic formulations, especially after recent advancements in entropy frameworks, such as the incorporation of quantum gravitational effects and beyond the extensivity of classical BG entropy. Following a similar procedure as described in Ref.~\cite{Bhattacharya:2021lgk}, one can obtain the general solution of Einstein field equation \eqref{1WTL1}  by using the relation $T_{r}^{r}=-\varepsilon/8\pi$ which yields
\begin{eqnarray}\label{2WTL4}
 f(r)&=&1-\frac{r_{\rm e}}{r}-\frac{1}{r}\int\limits_{r_{\rm e}}^{r}d\mathfrak{x}\ \mathfrak{x}^{2}\varepsilon(
 \mathfrak{x})=1-\frac{1}{r}\int\limits^{r}\mathfrak{x}^{2}\varepsilon(\mathfrak{x})d\mathfrak{x}-\frac{r_{\rm e}}{r}\left[1-\frac{1}{r_{\rm e}}\int\limits^{r_{\rm e}}\varepsilon(\mathfrak{x})\mathfrak{x}^{2}d\mathfrak{x}\right]~,
\end{eqnarray}
that gives
\begin{eqnarray}\label{2WTL5}
 f(r)&=&-r^{-1}\mathbb{C}+g(r)\,,
\end{eqnarray}
where $\mathbb{C}$ represents the integration constant and 
\begin{eqnarray}\label{2WTL6}
 g(\mathfrak{y})&=&1-\frac{1}{\mathfrak{y}}\int\limits^{\mathfrak{y}}\varepsilon(\mathfrak{x})\mathfrak{x}^{2}d\mathfrak{x}\,,
\end{eqnarray}
We identify the integration constant $\mathbb{C}$ with the mass of the BH due to the following considerations. According to the Birkhoff theorem given in Ref.~\cite{Birkhoff}, the external solution of SSS can be interpreted from the Schwarzschild metric in the absence of any matter. It is easy to obtain $g(r)=1$ and $f(r_{\rm e})=-r^{-1}\mathbb{C}+1$ for $\varepsilon(r)=0$ by utilizing Eq.~\eqref{2WTL6} and the integration constant $\mathbb{C}$ takes the following shape
\begin{eqnarray}\label{2WTL7}
\mathbb{C}&=&2M=r_{\rm e} \ g(r_{\rm e})\,,
\end{eqnarray}
and thus, we can compute 
\begin{eqnarray}\label{2WTL8}
f(r)&=&-2Mr^{-1}+g(r)~.
\end{eqnarray}
As we discussed earlier, the Einstein field equation for SSS provided in \eqref{1WTL1} is the differential equation of first order. Subsequently, solution \eqref{2WTL8} includes one integration constant, which indicates the mass of the BH. Additionally, the spacetime metric may incorporate parameters such as charge $Q$ obtained from the energy-momentum tensor $\varepsilon$; for example, in the RN case, we have $\varepsilon(r) =\frac{Q^{2}}{r^{4}}$. Therefore, one can express $g(r)=g(r,\nu^{i})$, where $\nu^{i}$ denotes the set of parameters (apart from mass) that enter the spacetime geometry, and the number of these parameters is assumed to be arbitrary. Moreover, we can denote the metric function as $f(r,~M,~\nu^{i})$. 
\subsection{Computing the Generalized Thermodynamic First Law}
Let us assume a small virtual adjustment to the parameters, resulting in a corresponding slight change in the horizon radius $\delta r_{\rm e}$. Here, a virtual variation indicates a comparison between nearby BH solutions rather than an actual dynamical process. We simply compare two BH solutions: one with horizon radius $r_{\rm e}$ (with mass $M$ and so forth) and one with virtual displacement $r_{\rm e}+\delta r_{\rm e}$  (with virtual displacement mass $M+\delta M$ and so forth). Thereby, by utilizing Eq.~\eqref{2WTL7}, we determine
\begin{eqnarray}\label{GFL}
2\delta M&=&\left[g(r_{\rm e},~\nu^{i})+\frac{\partial g(r_{\rm e},~\nu^{i})}{\partial r_{\rm e}}\ r_{\rm e}\right]\delta r_{\rm e}+r_{\rm e}\frac{\partial g(r_{\rm e},~\nu^{i})}{\partial \nu^{i}}\delta\nu^{i}\,.
\end{eqnarray}
By utilizing Eq.~\eqref{2WTL5}, we get the following relation
\begin{eqnarray}\label{GFL1}
f'(r_{\rm e})=\frac{\partial f(r)}{\partial r}\bigg|_{r=r_{\rm e}}&=&\frac{\partial g(r_{\rm e},~\nu^{i})}{\partial r_{\rm e}} +\frac{ g(r_{\rm e},~\nu^{i})}{ r_{\rm e}}~.
\end{eqnarray}
Thereby, one can obtain the explicit expression by employing Eq.~\eqref{2WTL8}
\begin{eqnarray}\label{GFL2}
\frac{\partial g(r_{\rm e},~\nu^{i})}{\partial \nu^{i}}&=&\frac{\partial f(r_{\rm e},~M,~\nu^{i})}{\partial \nu^{i}}\bigg|_{r=r_{\rm e},~M=0}~.
\end{eqnarray}
Now, we determine the explicit expression for the FL of BH thermodynamics by plugging Eqs.~\eqref{GFL1} and \eqref{GFL2} into Eq.~\eqref{GFL}, which is given as
\begin{eqnarray}\label{GFLT}
T\delta \mathcal{S}_{\rm HB}&=&\delta M -\frac{r_{\rm e}}{2}\frac{\partial f(r_{\rm e},~M,~\nu^{i})}{\partial \nu^{i}}\bigg|_{r=r_{\rm e},~M=0} \ \delta\nu^{i}~,
\end{eqnarray}
and after rearranging this equation, we get
\begin{eqnarray}\label{GFLT}
\delta M &=&T\delta \mathcal{S}_{\rm HB}+\frac{r_{\rm e}}{2}\frac{\partial f(r_{\rm e},~M,~\nu^{i})}{\partial \nu^{i}}\bigg|_{r=r_{\rm e},~M=0} \ \delta\nu^{i}~,
\end{eqnarray}
As we mentioned earlier, the variation in BH entropy is expressed solely through the changes in its parameters\footnotetext{Generally, the entropy of a BH is expressed in the form of its fundamental physical parameters, including mass, charge, and the cosmological constant. Any small change in these parameters can affect the BH's entropy without any external contribution. Therefore, the entropy variation in BHs inherently represents their thermodynamic nature, encoding the changes in horizon-related parameters.}. In our analysis, we do not need to define any thermodynamic potential here. In addition, the expression for temperature is $T=f'/4\pi$, and the BH entropy is one-fourth of the area of its event horizon, a widely accepted formula in the literature, as described in Ref.~\cite{Bhattacharya:2021lgk}.

Here, we discuss why we need to depart from the standard Hawking-Bekenstein entropy to other entropy frameworks. It is a well-known fact that the additive and extensive nature of BG thermodynamics and statistical mechanics \cite{Dabrowski:2024qkp} and the traditional BG entropy is generalized by the framework of non-extensive entropy \cite{Nojiri:2022aof,Nojiri:2022dkr,Nojiri:2022sfd,Elizalde:2025iku,Odintsov:2023vpj,Nojiri:2024zdu}. What differentiates Hawking-Bekenstein entropy is its linear relationship with the BH’s event horizon area, whereas, in conventional thermodynamics, entropy typically modifies with volume. Generalized entropy frameworks, as given in Ref.~\cite{Dabrowski:2024qkp}, are employed in BH thermodynamics to go beyond the traditional BG formulation, thereby allowing the inclusion of phenomena such as non-extensivity, strong correlations, and quantum-gravity effects. In this context, we now extend Eq.~\eqref{GFLT} to the generalized entropy framework. We begin our analysis by using the generalized form of four-parameter entropy as given in Refs.~\cite{Nojiri:2022aof,Nojiri:2022dkr,Nojiri:2022sfd,Elizalde:2025iku,Odintsov:2023vpj,Nojiri:2024zdu,Odintsov:2022qnn,Dabrowski:2024qkp}, which can be written as 
\begin{eqnarray}\label{GEF}
\mathcal{S}_\mathrm{GE}(\alpha_{\pm},~
\beta,~\sigma)&=&\frac{1}{\sigma}\left[\left(1+\frac{\alpha_{+}\mathcal{S}_{\rm HB}}{\beta}\right)^{\beta}-\left(1+\frac{\alpha_{-}\mathcal{S}_{\rm HB}}{\beta}\right)^{-\beta}\right]~,
\end{eqnarray}
where $\alpha_{\pm},~\beta$, and $\sigma$ are parameters that govern the generalized entropy framework, and for specific values of these parameters, one can obtain different non-extensive entropy frameworks. Similarly, there is a five-parameter entropy, which can be written as
\begin{eqnarray}\label{GEFF}
\mathcal{S}_\mathrm{GE}(\alpha_{\pm},~\
\beta,~\sigma,~\epsilon)&=&\frac{1}{\sigma} \left[ \left\{1 + \frac{1}{\epsilon} \tanh\left( \frac{\epsilon \alpha_{+} \mathcal{S}_{\rm HB}}{\beta} \right) \right\}^{\beta} - \left\{1 + \frac{1}{\epsilon} \tanh\left( \frac{\epsilon \alpha_{-} \mathcal{S}_{\rm HB}}{\beta} \right) \right\}^{-\beta} \right]~,
\end{eqnarray}
where $\alpha_{\pm},~\beta,~\sigma$ are the same parameters as those defined in Eq.~\eqref{GEF}. In the five-parameter generalized entropy, we have an additional parameter $\epsilon$. The values of these parameters must be positive, and $\mathcal{S}_{\rm HB}$ represents the Bekenstein-Hawking entropy. Let us mention here that if $\epsilon\to 0$ then Eq.~\eqref{GEFF} can be reduced to Eq.~\eqref{GEF} by making some adjustments as mentioned in Refs.~\cite{Nojiri:2022aof,Nojiri:2022dkr,Nojiri:2022sfd,Elizalde:2025iku,Odintsov:2023vpj,Nojiri:2024zdu,Odintsov:2022qnn,Dabrowski:2024qkp}\footnote{making some adjustments means that if we take $\epsilon\to0$ and by using Taylor series which gives $\lim\limits_{\epsilon \to 0}\frac{1}{\epsilon} \tanh\left( \frac{\epsilon \alpha_{-} \mathcal{S}_{\rm HB}}{\beta} \right)\to \frac{\alpha_{\pm}\mathcal{S}_{\rm HB}}{\beta}$. So, the relation mentioned in Eq.~\eqref{GEFF} takes the form $\lim\limits_{\epsilon \to 0}\left\{1 + \frac{1}{\epsilon} \tanh\left( \frac{\epsilon \alpha_{\pm} \mathcal{S}_{\rm HB}}{\beta} \right) \right\}=\left(1+\frac{\alpha_{\pm}\mathcal{S}_{\rm HB}}{\beta}\right)$.}. Similarly, we can define the generalized three parameter entropy $\mathcal{S}_{3}(\alpha_{+},~\beta,~\sigma)$ by putting $\epsilon\to 0,~\alpha_{-}\to0$ as provided in Refs.~\cite{Elizalde:2025iku,Nojiri:2022aof,Nojiri:2024zdu}. Thereby, mathematically, it can be expressed in the following form
\begin{eqnarray}\label{TSSME}
 \mathcal{S}_{3}(\alpha_{+},~\beta,~\sigma)=\frac{1}{\sigma}\left[\left(1+\frac{\alpha_{+} \mathcal{S}_{\rm HB}}{\beta}\right)^{\beta}-1\right]~,
\end{eqnarray}
which is also used to study various thermodynamic aspects of BHs. The entropy given in Eq.~\eqref{TSSME} follows the ab\'{e} addition rule, which we have provided in Appendix \ref{app:A}. The entropy deformation parameters are kept fixed throughout the thermodynamic variations considered below. The corresponding generalized FL can therefore be written as 

\begin{eqnarray}\label{GFLTGE}
\delta M &=&T_\mathrm{GE}\delta \mathcal{S}_\mathrm{GE}(\alpha_{\pm},~\
\beta,~\sigma) +\frac{r_{\rm e}}{2}\frac{\partial f(r_{\rm e},~M,~\nu^{i})}{\partial \nu^{i}}\bigg|_{r=r_{\rm e},~M=0} \ \delta\nu^{i}~.
\end{eqnarray}

The expression for the FL of BH thermodynamics in Eq.~\eqref{GFLTGE} applies to the class of spacetimes considered here without requiring specification of a particular entropy framework. Following the same procedure that we have adopted above, one can define the thermodynamic FL for generalized entropy with five parameters present in Eq. ~\eqref{GEFF}, which take the following shape
\begin{eqnarray}\label{GFLTGEF}
\delta M &=&T_\mathrm{GE}\delta \mathcal{S}_\mathrm{GE}(\alpha_{\pm},~\
\beta,~\sigma,~\epsilon) +\frac{r_{\rm e}}{2}\frac{\partial f(r_{\rm e},~M,~\nu^{i})}{\partial \nu^{i}}\bigg|_{r=r_{\rm e},~M=0} \ \delta\nu^{i}~.
\end{eqnarray}

Now, as we define the FL of BH thermodynamics for five- and four-parameter entropy, we can also define the FL of BH thermodynamics for generalized three-parameter entropy, which can be expressed as
\begin{eqnarray}\label{GFLTGEF}
\delta M &=&T_{\mathcal{S}_{3}}\delta \mathcal{S}_{3}(\alpha_{+},~\
\beta,~\sigma) +\frac{r_{\rm e}}{2}\frac{\partial f(r_{\rm e},~M,~\nu^{i})}{\partial \nu^{i}}\bigg|_{r=r_{\rm e},~M=0} \ \delta\nu^{i}~.
\end{eqnarray}
The three-parameter entropy given in Eq.~\eqref{TSSME} is algebraically simpler than the four- and five-parameter forms in Eqs.~\eqref{GEF} and \eqref{GEFF}.

\subsection{Generalized Smarr-type Formula}
Using Eq.~\eqref{2WTL7} for the BH's mass together with Eq.~\eqref{GFL1}, we obtain $g(r_{\rm e},~\nu^{i})$ which yields
\begin{eqnarray}\label{GSFT}
M&=&2T\mathcal{S}_{\rm HB}-\frac{r_{\rm e}^{2}}{2}\frac{\partial g(r_{\rm e}~\nu^{i})}{\partial r_{\rm e}}~,
\end{eqnarray}
where the expressions for entropy and temperature are already described in Eq.~\eqref{1WTL1}. Furthermore, the contribution from the second term in Eq.~\eqref{GSFT} arises in the presence of either the source term or the cosmological constant. For the Schwarzschild BH, where neither source nor the cosmological constant is present, we have $g(r_{\rm e}~\nu^{i})$, leading to the expression $M=2T\mathcal{S}_{\rm HB}$, which is the Smarr formula for this particular case (which is the Schwarzschild BH). Thereby, by using the relation given in Eq.~\eqref{2WTL8}, one can define
\begin{eqnarray}\label{GSFT1}
 g(r_{\rm e},~\nu^{i})=f(r_{\rm e},~M=0,~\nu^{i})\,.
\end{eqnarray}
Furthermore, one can determine the Smarr formula by inserting Eq.~\eqref{GSFT1} into Eq.~\eqref{GSFT} and it takes the following form
\begin{eqnarray}\label{GSF}
 M=2T\mathcal{S}_{\rm HB}-\frac{r_{\rm e}^{2}}{2}\left(\frac{\partial f(r,~M,~\nu^{i})}{\partial r}\right)\bigg|_{r=r_{\rm e},~M=0}\,.
\end{eqnarray}
It follows that, in SSS, one can extract the Smarr formula straightforwardly from Eq.~\eqref{GSF}, given the metric. Hence, Einstein's field equation yields not only the FL but also the Smarr-type relation. As mentioned earlier, we aim to construct the generalized Smarr-type relation for four-dimensional SSS without relying on the entropy framework, which can be used to study the thermodynamics of BHs. Therefore, by using the Eq.~\eqref{GEF}, one can obtain the following relation
\begin{eqnarray}\label{GSF1}
 T\mathcal{S}_{\rm HB}=T_\mathrm{GE}\mathcal{S}_{\rm HB}\left( \frac{\partial \mathcal{S}_\mathrm{GE}(\alpha_{\pm},~\beta,~\sigma)}{\partial \mathcal{S}_{\rm HB}}\right).
\end{eqnarray}
where 
\begin{eqnarray}\label{GSF2}
 \frac{\partial \mathcal{S}_\mathrm{GE}(\alpha_{\pm},~\beta,~\sigma)}{\partial \mathcal{S}_{\rm HB}}=\frac{1}{\sigma}\left[\alpha_{-}\left(1+\frac{\alpha_{-}\mathcal{S}_{\rm HB}}{\beta}\right)^{-1-\beta}+\alpha_{+}\left(1+\frac{\alpha_{+}\mathcal{S}_{\rm HB}}{\beta}\right)^{-1+\beta}\right].
\end{eqnarray}

Hence, we obtain the generalized Smarr-type relation in terms of the generalized entropy with four parameters by employing Eq.~\eqref{GSF1} in Eq.~\eqref{GSF}, which yields
\begin{eqnarray}\label{GSFF}
 M=2T_\mathrm{GE}\mathcal{S}_{\rm HB} \left(\frac{\partial \mathcal{S}_\mathrm{GE}(\alpha_{\pm},~\beta,~\sigma)}{\partial \mathcal{S}_{\rm HB}}\right)-\frac{r_{\rm e}^{2}}{2}\left(\frac{\partial f(r,~M,~\nu^{i})}{\partial r}\right)\bigg|_{r=r_{\rm e},~M=0}.
\end{eqnarray}

The Smarr relation we present in Eq.~\eqref{GSFF} applies to the class of four-dimensional SSS without assuming a particular entropy framework. Similarly, for the generalized Smarr relation in terms of generalized entropy with five parameters, we have the following relation, which is given as
\begin{eqnarray}\label{GSFF1}
 T\mathcal{S}_{\rm HB}=T_\mathrm{GE}  \mathcal{S}_{\rm HB}\left(\frac{\partial \mathcal{S}_\mathrm{GE}(\alpha_{\pm},~\
\beta,~\sigma,~\epsilon)}{\partial \mathcal{S}_{\rm HB}}\right),
\end{eqnarray}
where in the case of five-parameter entropy, we have
\begin{eqnarray}\nonumber
 \frac{\partial \mathcal{S}_\mathrm{GE}(\alpha_{\pm},~\beta,~\sigma,~\epsilon)}{\partial \mathcal{S}_{\rm HB}}&=&\frac{1}{\sigma }\bigg[\alpha_{+}\bigg\{1+ \frac{1}{\epsilon}\tanh \left(\frac{\alpha_{+} \mathcal{\mathcal{S}_{\rm HB}} \epsilon }{\beta}\right)^{\beta -1}\bigg\} \text{sech}^2\left(\frac{\alpha_{+} \mathcal{\mathcal{S}_{\rm HB}} \epsilon }{\beta}\right)+\alpha_{-}\bigg\{1+\frac{1}{\epsilon}\tanh \left(\frac{\alpha_{-} \mathcal{\mathcal{S}_{\rm HB}} \epsilon }{\beta}\right)^{-\beta -1}\bigg\} \\\label{FPGSF1}&\times&\text{sech}^2\left(\frac{\alpha_{-} \mathcal{\mathcal{S}_{\rm HB}} \epsilon }{\beta}\right)\bigg]~.
\end{eqnarray}
Hence, our generalized Smarr-type relation for five-parameter entropy can be defined by inserting Eq.~\eqref{FPGSF1} into Eq.~\eqref{GSF}, which yields 
\begin{eqnarray}\label{FPGSF}
 M=2T_\mathrm{GE} \left(\frac{\partial \mathcal{S}_\mathrm{GE}(\alpha_{\pm},~\
\beta,~\sigma,~\epsilon)}{\partial \mathcal{S}_{\rm HB}}\right)\mathcal{S}_{\rm HB}-\frac{r_{\rm e}^{2}}{2}\left(\frac{\partial f(r,~M,~\nu^{i})}{\partial r}\right)\Bigg|_{r=r_{\rm e},~M=0}~.
\end{eqnarray}

Moreover, by employing Eq.~\eqref{TSSME}, we compute the relation for generalized three-parameter entropy as follows
\begin{eqnarray}\label{FPGST}
 T\mathcal{S}_{\rm HB}=\frac{T_{\mathcal{S}_{3}} \ \beta  \left((\sigma  \mathcal{S}_{3}+1)^{1/\beta }\right)^{\beta -1} \left((\sigma  \mathcal{S}_{3}+1)^{1/\beta }-1\right)}{\sigma }~.
\end{eqnarray}
The generalized Smarr relation for the generalized three-parameter entropy can be derived by utilizing Eq.~\eqref{FPGST} in Eq.~\eqref{GSF}, which can be written as
\begin{eqnarray}\label{FPGST1}
 M=2 T_{\mathcal{S}_{3}} \left\{\frac{\beta  \left((\sigma  \mathcal{S}_{3}+1)^{1/\beta }\right)^{\beta -1} \left((\sigma  \mathcal{S}_{3}+1)^{1/\beta }-1\right)}{\sigma }\right\}-\frac{r_{\rm e}^{2}}{2}\left(\frac{\partial f(r,~M,~\nu^{i})}{\partial r}\right)\Bigg|_{r=r_{\rm e},~M=0}~~.
\end{eqnarray}
The generalized Smarr-type relation obtained above remains valid without imposing Euler homogeneity on the generalized entropy. For homogeneous $\mathcal{S}_{\rm GE}$ such as the power-law Tsallis-Cirto and Barrow entropies, Euler's theorem expresses the entropy term as one proportional to $T_{\rm GE}\mathcal{S}_{\rm GE}$. In the case of non-homogeneous entropy formulations, the deformation parameters are kept constant and the complete derivative contribution $2T_\mathrm{GE} \left(\frac{\partial \mathcal{S}_\mathrm{GE}(\alpha_{\pm},~\
\beta,~\sigma,~\epsilon)}{\partial \mathcal{S}_{\rm HB}}\right)\mathcal{S}_{\rm HB}$ must be retained.

It is also important to distinguish the Hawking temperature from the thermodynamic conjugate temperature arising within generalized entropy frameworks (as highlighted in Ref.~\cite{Nojiri:2021czz}). Hawking temperature arises from the surface gravity at the BH's event horizon, making it a geometric feature governed by the spacetime structure. On the other hand, the conjugate temperature arises from the thermodynamic definition $T_{\rm GE}=\partial{M}/\partial{\mathcal{S}_{\rm GE}}$ based on the selected entropy functional, and thereby reflects the microscopic (or statistical) structure of the entropy. The two coincide for the Hawking-Bekenstein entropy but generally differ for generalized entropy models. This distinction should be interpreted as arising from the modified thermodynamic framework of different entropy models, rather than indicating any violation of the semiclassical Hawking result. In this context, the Hawking temperature reflects the semiclassical behavior of the horizon, while the conjugate temperature describes the thermodynamic response of the BH when considering different entropy models.

Another interesting point is that applying different entropy models directly to the FL of BH thermodynamics, without changing the entropy-area relation, leads to inconsistencies, as reported in Refs.~\cite{Nojiri:2021czz,Zafar:2025nho}. To resolve this inconsistency, one first defines the thermodynamic mass $M(\mathcal{S})$ from the horizon condition for the chosen entropy model, then determines the temperature from the FL. This ensures that the temperature determined in this approach remains consistent with the generalized entropy, thereby addressing the inconsistency raised in Ref.~\cite{Nojiri:2021czz}. Furthermore, the generalized entropy function incorporates numerous well-known entropy models, such as Tsallis-Cirto, Barrow, R\'{e}nyi, Sharma-Mittal, Kaniadakis, and LQG-inspired entropies for appropriate parameter limits. We present the first law and Smarr-type relation for those entropies in Table \ref{Table-1}, while the explicit expressions and limiting procedures are given in Appendix \ref{app:A}.

\begin{table}[t]
\caption{\label{Table-1}
\raggedright We present a summary of the generalized FL of the BH thermodynamics and the generalized Smarr-type relation for different entropy models, derived from the generalized entropy framework. We also mention here that by employing these formalisms, one can derive the thermodynamic FL of BH and the Smarr relation for the BH solutions belonging to the class of SSS.}
\begin{ruledtabular}
\begin{tabular}{ccc}
$\text{Entropy Models}$ & FL $r=r_{\rm e}$&  The Smarr Relation at $r=r_{\rm e}$ \\
\hline
 Tsallis-Cirto  &$T_\mathbb{T}\delta \mathcal{S}_\mathbb{T}=\delta M -\frac{r_{\rm e}}{2}\frac{\partial f(r_{\rm e},~M=0,~\nu^{i})}{\partial \nu^{i}} \ \delta\nu^{i}$ & \small$2\tilde{\delta} T_\mathbb{T} \mathcal{S}_\mathbb{T}-\frac{r_{\rm e}^{2}}{2}\left(\frac{\partial f(r,~M=0,~\nu^{i})}{\partial r}\right)$ \\
 Barrow   &\small $T_\mathbb{B}\delta \mathcal{S}_\mathbb{B}=\delta M -\frac{r_{\rm e}}{2}\frac{\partial f(r_{\rm e},~M=0,~\nu^{i})}{\partial \nu^{i}} \ \delta\nu^{i}$ &\small $(2+\Delta)T_\mathbb{B}\mathcal{S}_\mathbb{B}-\frac{r_{\rm e}^{2}}{2}\left(\frac{\partial f(r,~M=0,~\nu^{i})}{\partial r}\right)$ \\ 
 R\'{e}nyi  &\small$T_\mathbb{R}\delta \mathcal{S}_\mathbb{R}=\delta M -\frac{r_{\rm e}}{2}\frac{\partial f(r_{\rm e},~M=0,~\nu^{i})}{\partial \nu^{i}} \ \delta\nu^{i}$ &\small $\frac{2\ T_\mathbb{R}(1-e^{-\alpha \mathcal{S}_\mathbb{R}})}{\alpha}-\frac{r_{\rm e}^{2}}{2}\left(\frac{\partial f(r,~M=0,~\nu^{i})}{\partial r}\right)$  \\
 Sharma-Mittal  &\small$T_\mathbb{SM}\delta \mathcal{S}_\mathbb{SM}=\delta M -\frac{r_{\rm e}}{2}\frac{\partial f(r_{\rm e},~M=0,~\nu^{i})}{\partial \nu^{i}} \ \delta\nu^{i}$ &\small $\frac{2\ T_\mathbb{SM}\left(1+R\mathcal{S}_\mathbb{SM}\right)^{1-\tilde{\delta}/R}\left\{-1+(1+R\mathcal{S}_\mathbb{SM})^{\tilde{\delta}/R}\right\}}{\tilde{\delta}}-\frac{r_{\rm e}^{2}}{2}\left(\frac{\partial f(r,~M=0,~\nu^{i})}{\partial r}\right)$ \\
 Kaniadakis  &\small$T_\mathrm{K}\delta \mathcal{S}_\mathrm{K}=\delta M -\frac{r_{\rm e}}{2}\frac{\partial f(r_{\rm e},~M=0,~\nu^{i})}{\partial \nu^{i}} \ \delta\nu^{i}$ & \small$\frac{2\ T_\mathrm{K}\sqrt{1+\mathcal{K}^{2}\mathcal{S}_\mathrm{K}^{2}} \sinh^{-1}(\mathcal{K}\mathcal{S}_\mathrm{K})}{\mathcal{K}}-\frac{r_{\rm e}^{2}}{2}\left(\frac{\partial f(r,~M=0,~\nu^{i})}{\partial r}\right)$ \\
 LQG & \small$T_\mathrm{L}\delta \mathcal{S}_\mathrm{L}=\delta M -\frac{r_{\rm e}}{2}\frac{\partial f(r_{\rm e},~M=0,~\nu^{i})}{\partial \nu^{i}} \ \delta\nu^{i}$ &{\small$\frac{2\ T_\mathrm{L}(1+\mathcal{S}_\mathrm{L}(1-q))\ln{\left(1+(1-q)\mathcal{S}_\mathrm{L}\right)}}{1-q}-\frac{r_{\rm e}^{2}}{2}\left(\frac{\partial f(r,~M=0,~\nu^{i})}{\partial r}\right)$}  \\
\hline
\end{tabular}
\end{ruledtabular}
\end{table}
\section{Ruppeiner Interpretation in the Generalized Entropy framework}\label{sec-4}
In this section, we discuss how generalized entropy modifies geothermodynamics formalisms, specifically the Ruppeiner metric of the RN BH. First, we discuss some basic background of the geothermodynamic formalism and its role in BH thermodynamics. In thermodynamics, differential geometry has traditionally been used to understand various thermodynamic phenomena by examining the geometry of the phase space. Phase transitions in non-extremal BHs are commonly analyzed using two complementary approaches: one based on the heat capacity \cite{Banerjee:2011cz,Banerjee:2012zm,Azreg-Ainou:2014gja,Mandal:2016anc,Majhi:2012fz,Ma:2014tka} and other on interpretation of cosmological constant  as thermodynamic pressure \cite{Banerjee:2016nse,Bhattacharya:2017hfj,Dehyadegari:2018pkb,Kubiznak:2012wp,Kubiznak:2016qmn,Majhi:2016txt,Bhattacharya:2017nru}. Furthermore, there are many ways to construct a thermodynamic geometry formalism; for example, one of the pioneering metric formalisms is the Weinhold metric, whose components are defined by the Hessian of the internal energy \cite{weinhold1975metric}. Subsequently, Ruppeiner \cite{Ruppeiner:1979bcp,ruppeiner1995riemannian,Ruppeiner:2008kd,Ruppeiner:2013yca,Bhattacharya:2019awq} extended this metric formalism of Weinhold by introducing a conformal factor in the form of the inverse of the temperature. 

By defining the generalized entropy as $\mathcal{S}_{\rm GE}=F(\mathcal{S}_{\rm HB})$, where $F$ is a differentiable entropy function, we derive the Ruppeiner curvature in terms of $F$ and its derivative and determine the class of entropy transformation that maintains the thermodynamic flatness of RN BH. Although thermodynamic curvature singularities are traditionally interpreted as signatures of critical phenomena, their relation to divergence in the response function is sensitive to the selection of the thermodynamic metric, potential, and ensemble. Moreover, the resulting curvature describes the thermodynamic geometry associated with the chosen entropy representation. Thus, its microscopic significance can only be understood within a suitable statistical description and does not necessarily provide a direct measure of microscopic interactions in BHs.

Now, we first discuss the metric function for RN BH in the spherically symmetric case, which is given as 
\begin{eqnarray}\label{LMEF}
ds^{2}=-\mathcal{U}(r)dt^{2}+\frac{1}{\mathcal{U}(r)}dr^{2}+r^{2}(d\theta^{2}+\sin\!{\theta}^{2}d\phi^{2})\,, \quad \text{where}~~~~ \mathcal{U}(r)=1-\frac{2M}{r}+\frac{\mathcal{Q}^{2}}{r^{2}}\,.
\end{eqnarray}
One can easily express the inner and outer horizons for an RN BH given as
\begin{eqnarray}\label{root}
r_{\pm}=M\pm\sqrt{M^{2}-\mathcal{Q}^{2}}\,.
\end{eqnarray}
Extremality is characterized by $M^{2}=Q^{2}$ for which the inner and outer horizons coincide $r_{-}=r_{+}=M=|Q|$. It is therefore convenient to define 
\begin{eqnarray}\label{assu}
x\equiv \mathcal{S}_{\rm HB}=\pi\ r_{+}^{2}\,,\text{and}~~ \mathcal{S}_{\rm HB}=\pi \left(M+\sqrt{M^{2}-\mathcal{Q}^{2}}\right)^{2},
\end{eqnarray}
and this relation given in Eq.~\eqref{root} can be expressed the generalized entropy as a differentiable function of the Hawking-Bekenstein entropy, as follows
\begin{eqnarray}\label{assu}
\mathcal{S}_{GE}=F(x)\,, \quad x=\mathcal{S}_{\rm HB}
\end{eqnarray}
by utilizing Eq.~\eqref{assu}, one can express the Ruppeiner metric in the two-dimensional thermodynamic state space $X^{i}=(M, Q)$, which can be written as 
\begin{eqnarray}\label{assu1}
g^\mathrm{Rup}_{ij}=-\frac{\partial^{2}{F(x)}}{\partial{X_{i}}\partial{X_{j}}}\,,
\end{eqnarray}
where $'$ denotes the derivative with respect to x. By using the chain rule, one can write the above equation in the following form
\begin{eqnarray}\label{assu2}
g^\mathrm{Rup}_{ij}=-\left(F''(x)\frac{\partial x}{\partial X^{i}}\frac{\partial x}{\partial X^{j}}+F'(x)\frac{\partial^{2}x}{\partial X^{i}\partial X^{j}}\right)\,.
\end{eqnarray}

Now, it is convenient to obtain the Ruppeiner curvature from the above metric, which is given as 
\begin{eqnarray}\label{assu2}
R^{\rm Rup}=\frac{2 x F''(x) \left\{F'(x) \left(x F'''(x)+F''(x)\right)-x F''(x)^2\right\}}{F'(x)^2 \left\{2 x F''(x)+F'(x)\right\}^2}\,. 
\end{eqnarray}
In the non-extremal state space $M^{2}>Q^{2}$ the entropy dependent regularity and non-degeneracy conditions are
\begin{eqnarray}\label{cft}
    F'(x)\neq0\,,\quad 2 x F''(x)+F'(x)\neq0\,.
\end{eqnarray}

This equation demonstrates the Ruppeiner curvature scalar for a generalized differentiable function $F$, while the corresponding thermodynamic metric is regular and nondegenerate only when the necessary conditions are satisfied. Therefore, if either condition fails, the Hessian metric becomes degenerate, rather than describing the curvature singularity of a regular thermodynamic manifold. Moreover, we can also rewrite the above equation as follows
\begin{eqnarray}\label{assu2}
R^{\rm Rup}=\frac{2 x F''(x)}{\left\{2 x F''(x)+F'(x)\right\}^2}\times\frac{d}{dx}\left(\frac{xF''(x)}{F'(x)}\right) \,, 
\end{eqnarray}

Equation \eqref{assu2} shows that if the quantity $\left(\frac{xF''(x)}{F'(x)}\right)=c$ (constant), then the Ruppeiner metric for the RN BH becomes thermodynamically flat. Furthermore, if we integrate this relation, then it gives 
\begin{eqnarray}\label{assu3}
F(x)=\mathcal{A}x^{\tilde{\gamma}}+\mathcal{B}, ~\text{where}~ \quad~\tilde{\gamma}=c+1,~~ \text{and}~~ c\neq -1 \,, 
\end{eqnarray}
where $\mathcal{A}\neq0$ and $\mathcal{B}$ are the integration constants\footnote{One can obtain different entropy models by fixing the normalization constant $\mathcal{A}$, $\mathcal{B}$ and $\tilde{\gamma}$. For example, if we put $\mathcal{A}=\tilde{\gamma}=1,~\mathcal{B}=0$, then Eq.~\eqref{assu3} reduces to Bekenstein-Hawking entropy; if we put $\tilde{\gamma}=\tilde{\delta}$ and  $\tilde{\gamma}=1+\Delta/2$, then we can easily obtain Tsallis-Cirto and Barrow entropy, respectively. Furthermore, another interesting point is that if we look at Eq.~\eqref{c32}, one can observe that it resembles the logarithmic structure of R\'{e}yni entropy. However, the logarithmic branch is not equivalent to R\'{e}nyi entropy; for example, if we set $\mathcal{A}=1/\alpha$ and $\mathcal{B}=1$, we still do not obtain R\'{e}nyi entropy. Furthermore, another reason is that if we consider $F(x)=1/\alpha\ln(1+\alpha x)$, then by putting this function into $\left(\frac{xF''(x)}{F'(x)}\right)\neq c$, it does not satisfy the flatness condition, as is also evident from Table \ref{Table-2}. }. If we consider $c=-1$, the relation must be integrated separately, which gives the logarithmic solution expressed as 
\begin{eqnarray}\label{c32}
F(x)=\mathcal{A}\ln{x}+\mathcal{B}\,. 
\end{eqnarray}

By applying homogeneous power-law modifications of the Bekenstein-Hawking entropy, a vanishing thermodynamic curvature is maintained across the regular thermodynamic state space. The Bekenstein-Hawking, Tsallis-Cirto, and Barrow entropy models fall within the flat class characterized by Eq.~\eqref{assu3}, except at $\tilde{\gamma}=1/2$, where $2 x F''(x)+F'(x)=0$ leads to metric degeneracy rather than a regular flat manifold. Now, we take the generalized three-parameter entropy as
\begin{eqnarray}\label{gtpef}
    F(x)=\frac{1}{\sigma }\left\{\left(\frac{ \alpha_{+} x}{\beta }+1\right)^{\beta }-1\right\}\,.
\end{eqnarray}

Utilizing Eq.~\eqref{gtpef} and Eq.~\eqref{assu2}, gives us the following expression 
\begin{eqnarray}\label{gtpef1}
    R^{\rm  Rup}=\frac{2 \alpha _+ (\beta -1)^2 \sigma  x \left(\frac{\alpha _+ x}{\beta }+1\right)^{-\beta }}{\left(\beta +\alpha _+ (2 \beta -1) x\right)^2}\,,
\end{eqnarray}
and taking the extremal limit $M^{2}-Q^{2}\to 0^{+}\implies x\to x_{Ext}=\pi Q^{2}$, it yields
\begin{eqnarray}\label{gtpef2}
 R^{\rm  Rup}_{\rm Ext}=\frac{2 \pi  \alpha _+ (\beta -1)^2 Q^2 \sigma  \left(\frac{\pi  \alpha _+ Q^2}{\beta }+1\right)^{-\beta }}{\left(\beta +\pi  \alpha _+ (2 \beta -1) Q^2\right)^2}\,.
\end{eqnarray}

Equation \eqref{gtpef2} is a generalized expression for RN BH, but it can be reduced to other entropy models for specific parameter choices discussed in Appendix \ref{app:A}, except for the Kaniadakis entropy.  However, for the case of the singular parameter, it is suitable to evaluate Eq.~\eqref{assu2} from the associated entropy functional prior to taking the extremal limit. Table \ref{Table-2} demonstrates that thermodynamic geometry for RN BH is governed by the underlying entropy representation. The flat Ruppeiner geometry for Bekenstein-Hawking entropy remains flat throughout the homogeneous power-law class discussed in Eq.~\eqref{assu3} (which also agrees with the result given in Ref.~\cite{Aman:2005xk,Aman:2003ug,Aman:2006kd}), including the Tsallis-Cirto and Barrow entropy models. On the other hand, the particular non-power-law entropy deformations, including the R\'{e}nyi, Sharma-Mittal, Kaniadakis, and LQG entropies, generically yield finite non-vanishing curvature in the regular thermodynamic regime. These curvature contributions can be reduced to vanishing curvature by using the undeformed limits discussed in Table \ref{Table-2}. Furthermore, we can also notice from Eq.~\eqref{assu3} that curvature divergences are possible whenever $(F'(x)+2xF''(x)=0)$ and the numerator remains finite and nonzero, indicating that these singularities result from the degeneracy of the thermodynamic Hessian rather than the extremality condition $M^{2}=Q^{2}\nRightarrow R^{\rm Rup}\to\infty$. 
\begin{table}[t]
\caption{\label{Table-2}
\raggedright In this table, we present $R^{\rm Rup}$ of the RN BH in the extremal limit for different entropy models, evaluated by taking $M^{2}-Q^{2}\to0^{+}$. Here, flat family means the solution of $\frac{xF''(x)}{F'(x)}=c$, which we present in Eq.~\eqref{assu3}. }
\begin{ruledtabular}
\begin{tabular}{ccccc}
$\text{Entropy Models}$ & Flat family& Function form&$R^\mathrm{Rup}$ & Limits reduce \\& & $F(x)=$&extremal limit & to undeformed $R^{\rm Rup}$
\\
\hline
Hawking-Bekenstein & \checkmark & $x$ & zero&--\\
 Tsallis-Cirto  & \checkmark & $x^{\tilde{\delta}}$ & zero&--\\
 Barrow   & \checkmark & ~$x^{1+\Delta/2}$ &zero&--\\
 R\'{e}nyi  & $\times$ & $\frac{1}{\alpha}\ln\left(1+\alpha x\right)$ & $\frac{2 \pi  \alpha ^2 Q^2}{\left(\pi  \alpha  Q^2-1\right)^2}$&$\alpha\to 0$\\
 Sharma-Mittal  & $\times$ & $\frac{1}{R}\left[\left(1+\tilde{\delta} x\right)^{R/\tilde{\delta}}-1\right]$ & $\frac{2 \pi  Q^2 (R-\tilde{\delta} )^2 \left\{\pi  \tilde{\delta}  Q^2+1\right\}^{-\frac{R}{\tilde{\delta} }}}{\left\{\pi  Q^2 (2 R-\tilde{\delta} )+1\right\}^2}$&$R,\tilde{\delta}\to 0$\\
 Kaniadakis  & $\times$ & $\frac{Sinh[\mathcal{K} x]}{\mathcal{K}}$ & $\frac{2 \pi  K^2 Q^2 \left\{2 \pi  K Q^2 \text{csch}\left(2 \pi  K Q^2\right)+1\right\} \text{sech}\left(\pi  K Q^2\right)}{\left\{2 \pi  K Q^2+\coth \left(\pi  K Q^2\right)\right\}^2}$&$\mathcal{K}\to 0$\\
 LQG-inspired exponential & $\times$ & $\frac{e^{x (1-q)}-1}{1-q}$ & $\frac{2 \pi  Q^2 (q-1)^2 e^{\pi  M^2 (q-1)}}{\left\{1-2 \pi  Q^2 (q-1)\right\}^2}$&$q\to 1$\\
\end{tabular}
\end{ruledtabular}
\end{table}
\section{Conclusions}\label{sec-6}
In this work, we explored generalized entropy models involving three-, four-, and five-parameter deformations and developed a consistent, unified geometric approach to BH thermodynamics. The primary goal of our study was to establish a generalized FL that holds for the class of static, spherically symmetric spacetimes considered here, regardless of the chosen non-additive and non-extensive entropy framework. In contrast to approaches that impose a modified entropy phenomenologically, we derived the generalized FL directly from the horizon thermodynamic relation of the Einstein equations, thereby ensuring consistency at both the dynamical and geometrical levels. In addition, we derived a generalized Smarr-type relation that holds for any entropy framework and for the static SSS configurations considered here. The resulting general Smarr-type relation does not require Euler homogeneity of the generalized entropy. Importantly, the generalized Smarr-type relation distinguishes entropy deformations that admit an Euler-homogeneous reduction from non-homogeneous entropy models for which the full entropy derivative must be retained.  

Furthermore, we analyzed the Ruppeiner geometry of the RN BH by using the generalized entropy formulations. A key result of our Ruppeiner thermodynamic geometry analysis is the establishment of a structural criterion that identifies whether a generalized entropy preserves or modifies the flat Ruppeiner geometry of the RN BH. By introducing the generalized entropy as a function of Bekenstein-Hawking entropy, the generalized curvature formula shows that regular homogeneous power-law transformations, $(F(\mathcal{S})=\mathcal{S}^{\tilde{\gamma}}) $, together with the separate logarithmic branch, preserve the flat Ruppeiner curvature. This provides a unified explanation for the flat thermodynamic geometry corresponding to the Bekenstein-Hawking, Tsallis-Cirto, and Barrow entropies. However, R\'{e}nyi, Sharma-Mittal, Kaniadakis, and LQG entropies usually give a finite, non-vanishing thermodynamic curvature that disappears in the undeformed limit, signaling that the non-vanishing curvature is associated with the specific entropy deformation rather than with the extremal condition, which merely determines the degeneracy (merging) of the inner and outer horizons. We also observed that entropy composability provides a useful statistical basis for classification, but on its own, it cannot determine whether the Ruppeiner geometry is flat, curved, or singular. This distinction among geometric extremality, entropy deformation, and thermodynamic curvature establishes a general framework for evaluating generalized entropy models in BH thermodynamics. The generalized framework proposed in our work can be further applied to a broad range of BH configurations, such as AdS, higher-dimensional, additional fields (such as quintessence, perfect fluid, and string clouds), and BH solutions in higher-order curvature gravities, where the interaction between generalized entropy and spacetime dynamics could reveal novel thermodynamic features.
\section*{Acknowledgments}

Usman Zafar acknowledges the Japanese government (MEXT) scholarship. This work of K. Bhattacharya is supported partly by the New Faculty Seed Grant (NFSG) of BITS Pilani, Dubai Campus. Moreover, the work of Kazuharu Bamba was supported in part by the JSPS KAKENHI
Grants No. 24KF0100 and No. 25KF0176, and a grant-in-aid of academic
research of the Yamaguchi Scholarship Foundation. 

\appendix
\section{Different Entropy Models obtained from Generalized Entropy\label{app:A}}
In this Appendix, we extend our analysis to study the implications of our thermodynamic FL and the Smarr relation beyond extensivity and additivity by choosing parameter values to obtain various entropy models, as described in Ref.~\cite{Dabrowski:2024qkp}. We employ Eq.~\eqref{GEFF} to derive different entropy formalisms as discussed in Refs.~\cite{Nojiri:2024zdu}. Firstly, if $\epsilon\to 0$, we obtain the four-parameter entropy given in Eq.~\eqref{GEF}, and then specify the remaining parameter values for different entropy models. 

If we consider $\alpha_{+}=\infty$ and $\alpha_{-}=0$ then Eq.~\eqref{GEF} reduces to the given form, which is written as 
\begin{eqnarray}\label{GEFAP}
 \mathcal{S}_\mathrm{GE}=\frac{1}{\sigma}\left(\frac{\alpha_{+}}{\beta}\right)^{\beta}\mathcal{S}_{\rm HB}^{\beta}~.
\end{eqnarray}
Equation \eqref{GEFAP} can be further simplified by choosing $\sigma=\left(\frac{\alpha_{+}}{\beta}\right)^{\beta}$, which gives
\begin{eqnarray}\label{TSBE}
 \mathcal{S}_\mathrm{GE}=\mathcal{S}_{\rm HB}^{\beta}~.
\end{eqnarray}
One can obtain Tsallis-Cirto entropy by choosing $\beta=\tilde{\delta}$ as given in Ref.~\cite{Tsallis:1987eu}, and for Barrow entropy, we can choose $\beta=1+\Delta/2$ \cite{Barrow:2020tzx}. A fascinating aspect is that it maps straightforwardly onto Barrow entropy by adopting the relation $\tilde{\delta}=1+\Delta/2$, where $\Delta$ impacts the degree of non-extensivity and non-additivity indirectly, and the permissible range for $\Delta\in[0,~1]$ corresponds to $\tilde{\delta}\in[1,~3/2]$. One thing common to both of these entropy models is that, by substituting $\Delta=0$ (Barrow entropy case) and $\tilde{\delta}=1$ (Tsallis-Cirto), one can recover the Hawking-Bekenstein entropy. Furthermore, if we take  $\epsilon\to 0,~\alpha_{-}\to0,~\beta\to0$ and $\frac{\alpha_{+}}{\beta}\to \text{finite}$ as described in Ref.~\cite{Nojiri:2024zdu}, then Eq.~\eqref{GEFF} reduces to the following form
\begin{eqnarray}\label{TSRE}
 \mathcal{S}_\mathrm{GE}=\frac{1}{\sigma}\left[\exp\left\{ \beta\ln\left(1+\frac{\alpha_{+}\mathcal{S}_{\rm HB}}{\beta}\right)\right\}-1\right]\approx~\frac{\beta}{\sigma}\ln\left(1+\frac{\alpha_{+}\mathcal{S}_{\rm HB}}{\beta}\right)~,
\end{eqnarray}
and by considering that $\sigma=\alpha_{+}$ and $\frac{\alpha_{+}}{\beta}=\alpha$, we obtain R\'{e}nyi entropy as given in Refs.~\cite{Odintsov:2022qnn,Rényi}. Its explicit form is presented as 
\begin{eqnarray}\label{TSREE}
 \mathcal{S}_\mathrm{GE}=\mathcal{S}_\mathbb{R}=\frac{1}{\alpha}\ln\left(1+\alpha \mathcal{S}_{\rm HB}\right)~.
\end{eqnarray}
The R\'{e}nyi entropy plays an important role in the generalized entropy frameworks owing to its additive property and its mathematical association with Tsallis-Cirto entropy through logarithmic correspondence. In the case of the Sharma-Mittal entropy \cite{SayahianJahromi:2018irq}, taking $\epsilon\to0$ and $ \alpha_{-}\to0$ reduces the generalized entropy in Eq.~\eqref{GEFF} to the reduced form in Eq.~\eqref{TSSME}. By using $\sigma=R,~\alpha_{+}=R$ and $\beta=R/\tilde{\delta}$, it is easy to obtain the exact form of the Sharma-Mittal entropy as given in Refs.~\cite{Odintsov:2022qnn,Dabrowski:2024qkp,SayahianJahromi:2018irq}, which takes the following shape
\begin{eqnarray}\label{TSSMEE}
 \mathcal{S}_\mathrm{GE}=\mathcal{S}_\mathbb{SM}=\frac{1}{R}\left[\left(1+\tilde{\delta} \mathcal{S}_{\rm HB}\right)^{R/\tilde{\delta}}-1\right]~.
\end{eqnarray}

Equation \eqref{TSSMEE} shows that the Sharma-Mittal entropy incorporates two parameters, which are a generalized combination of the Tsallis-Cirto and R\'{e}nyi entropies, and it provides a unique framework to study thermodynamic systems that are neither additive nor extensive. Sharma-Mittal entropy provides a more flexible framework for studying the modified BH thermodynamics, particularly when fractal geometries or quantum corrections are incorporated. Moreover, if we take $\epsilon\to0,~\alpha_{+}=\alpha_{-}=\sigma/2=\mathcal{K}$ and $\beta\to\infty$ after straightforward algebraic manipulation, then we have
\begin{eqnarray}\label{TSKE}
 \mathcal{S}_\mathrm{GE}=\frac{1}{2\mathcal{K}}\lim\limits_{\beta\to\infty}\left[\left(1+\frac{\mathcal{K}\mathcal{S}_{\rm HB}}{\beta}\right)^{\beta}-\left(1+\frac{\mathcal{K}\mathcal{S}_{\rm HB}}{\beta}\right)^{-\beta}\right]=\frac{1}{2\mathcal{K}}\left(e^{\mathcal{K}\mathcal{S}_{\rm HB}}-e^{-\mathcal{K}\mathcal{S}_{\rm HB}}\right)=\frac{1}{\mathcal{K}}\sinh{\mathcal{K}\mathcal{S}_{\rm HB}}~.
\end{eqnarray}
The above equation resembles the Kaniadakis entropy as given in Refs.~\cite{Kaniadakis:2002zz,Kaniadakis:2005zk,Drepanou:2021jiv}. The range for the $\mathcal{K}$-parameter is $\mathcal{K}\in (-1,1)$, and one can recover Hawking-Bekenstein entropy by setting $\mathcal{K}\to0$ \cite{Anand:2025cer,NooriGashti:2024ywc,Baruah:2024lcj}. Similarly, if we put $\epsilon\to0,~\alpha_{-}\to0,~\beta\to\infty$ and $\sigma=\alpha_{+}=1-\mathfrak{q}$, then Eq.~\eqref{GEFF} gives
\begin{eqnarray}\label{TSKE}
 \mathcal{S}_\mathrm{GE}=\frac{1}{\mathfrak{\sigma}}\left[\left(1+\frac{\alpha_{+}\mathcal{S}_{\rm HB}}{\beta}\right)^{\beta}-1\right]=\frac{1}{1-\mathfrak{q}}\lim\limits_{\beta\to\infty}\left[\left(1+\frac{(1-\mathfrak{q})\mathcal{S}_{\rm HB}}{\beta}\right)^{\beta}-1\right]=\frac{1}{1-\mathfrak{q}}[e^{(1-\mathfrak{q})\mathcal{S}_{\rm HB}}-1]~,
\end{eqnarray}
which is similar to the entropy given in Refs.~\cite{Majhi:2017zao,Liu:2021dvj}. It provides an entropy deformation with a non-linear exponential dependence on the Bekenstein-Hawking entropy. In addition, the entropy function given in Eq.~\eqref{GEFF} exhibits certain characteristics. For example, if $\mathcal{S}_{\rm HB}\to 0$, then $\mathcal{S}_{\rm GE}\to 0$, indicating that the entropy is normalized to vanish in the zero-entropy limit of the Hawking-Bekenstein description. Furthermore, the limits of the parameters summarized above yield the entropy models given in Table~\ref{Table-1}, along with their corresponding FL and Smarr-type relations.

\section{A Transformation of Ruppeiner Metric  in Adapted coordinates\label{app:B}}
In this Appendix, to clarify the generalized RN Ruppeiner geometry, we express the metric in coordinates associated with the Bekenstein-Hawking entropy. We construct the Ruppeiner metric first in the original coordinates $(M, Q)$, following the procedure presented in Sec. \ref{sec-4}, and the resulting metric is then expressed in terms of $(x, Q)$. This approach ensures the thermodynamic Hessian structure while accounting for the contribution arising from the entropy mapping $F(x)$. Incorporating another modification of variables eliminates the mixed terms, and the metric becomes diagonal, from which a compact analytical expression for the Ruppeiner curvature is derived.  Now, we transform the coordinates from $(M, Q)$ to $(x, Q)$ and by using the relation $\mathcal{U}(r_{+})=0$, we get 
\begin{eqnarray}\label{massss}
    M=\frac{r^{2}_{+}+Q^{2}}{2r_{+}}\,,\qquad \text{and} \qquad r_{+}=\sqrt{\frac{x}{\pi}}\,.
\end{eqnarray}
From this, we can determine
\begin{eqnarray}\label{massss1}
    M=\frac{x+\pi Q^{2}}{2\sqrt{\pi x}}\,,
\end{eqnarray}
Now, we can obtain the metric components after the metric transformation $(x, Q)$
\begin{eqnarray}\label{massss2}
   g_{xx}=-F''-F'\frac{x-3\pi Q^{2}}{2x(x-\pi Q^{2})}\,,~~~~~~g_{xQ}=g_{Qx}=-\frac{2\pi Q F'}{x-\pi Q^{2}}\,,~~~~ g_{QQ}=\frac{4\pi x F'}{x-\pi Q^{2}}\,.
\end{eqnarray}
The Hessian is not assumed to transform covariantly under a nonlinear change of variables; instead, we construct the Hessian first and then transform the resulting metric. Now, we do some more algebra to remove the mixed terms and make this metric diagonal, for which we consider
\begin{eqnarray}\label{massss3}
   a=\frac{Q}{\sqrt{x}}\,,\implies Q=a\sqrt{x}\,,
\end{eqnarray}
and from this, one obtains 
\begin{eqnarray}\label{massss4}
   dQ=\sqrt{x}da+\frac{Q}{2x}dx\,.
\end{eqnarray}

Now, by using the above relation, the cross terms are eliminated, and we obtain
\begin{eqnarray}\label{massss5}
   ds^{2}=-\frac{F'+2xF''}{2x}dx^{2}+\frac{4\pi F'x}{1-\pi a^{2}}da^{2}\,,
\end{eqnarray}
and we can define
\begin{eqnarray}\label{massss6}
   \chi=\frac{1}{\sqrt{\pi}} \sin^{-1}\!(\sqrt{\pi}a)\,,\quad \text{and}\qquad d\chi=\frac{da}{\sqrt{1-\pi a^{2}}}.
\end{eqnarray}
Finally, Eq.~\eqref{massss5} becomes
\begin{eqnarray}\label{massss7}
   ds^{2}=-\frac{F'+2xF''}{2x}dx^{2}+4\pi F'xd\chi^{2}\,.
\end{eqnarray}
This Eq.~\eqref{massss7} enables us to compute the Ricci curvature scalar given in Eq.~\eqref{assu2} for the Ruppeiner metric directly.

\end{document}